\documentclass[prl,twocolumn,showpacs,floatfix,superscriptaddress]{revtex4-1} 
\usepackage{amsmath, graphicx}
\usepackage{verbatim}  

\usepackage{color}

\newcommand{\exciting}{{\usefont{T1}{lmtt}{b}{n}exciting }}

\begin{document}

\title{
Mapping atomic orbitals with the transmission electron microscope: Images of defective graphene predicted from first-principles theory}

\author{Lorenzo Pardini}
\affiliation{Physics Department and IRIS Adlershof, Humboldt-Universit\"at zu Berlin, 12489 Berlin, Germany}
\author{Stefan L\"offler}
\affiliation{Institute of Solid State Physics, Vienna University of Technology, Austria}
\affiliation{University Service Centre for Transmission Electron Microscopy, Vienna University of Technology, Austria }
\affiliation{Department of Materials Science and Engineering, McMaster University, Hamilton, Ontario, Canada}
\author{Giulio Biddau}
\affiliation{Physics Department and IRIS Adlershof, Humboldt-Universit\"at zu Berlin, 12489 Berlin, Germany}
\author{Ralf Hambach}
\affiliation{Electron Microscopy Group of Materials Science, University of Ulm, 89081 Ulm, Germany}
\affiliation{Fraunhofer Institute for Applied Optics and precision engineering IOF, Jena, Germany}
\author{Ute Kaiser}
\affiliation{Electron Microscopy Group of Materials Science, University of Ulm, 89081 Ulm, Germany}
\author{Claudia Draxl}
\affiliation{Physics Department and IRIS Adlershof, Humboldt-Universit\"at zu Berlin, 12489 Berlin, Germany}
\affiliation{European Theoretical Spectroscopy Facility (ETSF)}
\author{Peter Schattschneider}
\affiliation{Institute of Solid State Physics, Vienna University of Technology, Austria}
\affiliation{University Service Centre for Transmission Electron Microscopy, Vienna University of Technology, Austria }

\begin{abstract}
Transmission electron microscopy has been a promising candidate for mapping atomic orbitals for a long time. Here, we explore its capabilities by a first principles approach. For the example of defected graphene, exhibiting either an isolated vacancy or a substitutional nitrogen atom, we show that three different kinds of images are to be expected, depending on the orbital character. To judge the feasibility of visualizing orbitals in a real microscope, the effect of the optics' aberrations is simulated. We demonstrate that, by making use of energy-filtering, it should indeed be possible to map atomic orbitals in a state-of-the-art transmission electron microscope.
\end{abstract}


\maketitle



The possibility to {\it see} atomic orbitals has always attracted great scientific interest. At the same time, however, the real meaning of "measuring orbitals" has been a subject that scientists have long and much dwelt upon (see, e.g., \cite{schwarz} and references therein). In the past, significant efforts have been devoted to the development of experimental approaches and theoretical models that allow for orbital reconstruction from experimental data \cite{schwarz}. Based on the generation of higher harmonics by femtoseconds laser pulses, a tomographic  reconstruction of the highest occupied molecular orbital (HOMO) for simple diatomic molecules in the gas phase was proposed \cite{itatani}. Direct imaging of the HOMO and the lowest-occupied molecular orbital (LUMO) of pentacene on a metallic substrate was theoretically predicted and experimentally verified with scanning-tunnelling microscopy (STM) \cite{PhysRevLett.94.026803}. More recently, real-space reconstruction of molecular orbitals from angle-resolved photoemission data has been demonstrated \cite{Puschnig30102009}. This method has been subsequently further developed to retrieve both the spatial distribution \cite{PhysRevB.84.235427} and the phase of electron wavefunctions of pentacene and perylene-3,4,9,10-tetracarboxylic dianhydride (PTCDA) adsorbed on silver \cite{Luftner16122013}. 

The reconstruction of charge densities and chemical bonds using transmission electron microscopy (TEM) has been considered \cite{Zuo1999, meyerexperimental2011, haruta2013, PhysRevB.88.115120, oxley2014}, but only recently the possibility of probing selected transitions to specific unoccupied orbitals by using energy-filtered TEM (EFTEM) was demonstrated theoretically. A first example for the capability of this approach was provided with the oxygen K-edge of rutile TiO$_2$ \cite{Loffler201339}. However the interpretation of experimental TEM images for systems like rutile would be complicated because of the multiple elastic scattering of electrons that occurs in thick samples.

In this Letter, we suggest defective graphene \cite{geim,PhysRevLett.93.187202,PhysRevB.70.245416,PhysRevLett.96.046806,PhysRevB.75.125408} as the prototypical two dimensional (2D) material to demonstrate the possibility of mapping atomic orbitals using EFTEM . We break the ideal $sp^2$ hybridization by introducing two different kinds of defects, namely a single isolated vacancy and a substitutional nitrogen atom. This lifts the degeneracy of the $p$-states, inducing strong modifications to the electronic properties compared to the pristine lattice \cite{PhysRevB.70.245416, novoselovtwo2005, sonhalf2006, heerschebipolar2007,PhysRevLett.95.205501,PhysRevB.87.165401, PhysRevB.62.7639, Ewels2002178}. 
By selecting certain scattering angles, dipole-allowed transitions dominate the electron energy loss spectroscopy (EELS) signal \cite{Jorissen}. A single-particle description can be safely adopted, since many-body effects do not play a major role in the excitation process. Overall, TEM images of these systems can be interpreted in terms of bare $s-p$ transitions. 

In an EFTEM experiment, an incoming beam of high-energy electrons (of the order of 100 keV) is directed to the target where it scatters at the atoms either elastically or inelastically. The outgoing electron beam is detected and analyzed. State of the art image simulations generally only include elastic scattering of the electrons using the multi-slice approach \cite{kirkland}. In the case of EFTEM for a thin sample, the influence of elastic scattering becomes negligible, and inelastic scattering gives the dominant contribution to the formation of the images.

The key quantity to describe the inelastic scattering of electrons, which is probed by EELS, is the mixed dynamic form factor (MDFF). It can be interpreted as a weighted sum of transition matrix elements between initial and final states $\phi_{i}$ and $\phi_{j}$ of the target electron \cite{Schattschneider2000333, Nelhiebel1999, Kohl1985}:

\begin{widetext}
\begin{equation}
 S\left( \bf{q}, \bf{q'}; \it{E} \right) = \sum_{i,j}\left\langle \phi_{i} \left| e^{i\bf{q} \cdot \bf{r}} \right| \phi_{j} \right\rangle \left\langle \phi_{j} \left| e^{-i\bf{q'} \cdot \bf{r}} \right| \phi_{i} \right\rangle \delta \left( E_j - E_i - E \right)
\end{equation}
\end{widetext}
with energies $E_i$ and $E_j$. $E$ is the energy-loss of the fast electron of the incident beam, $\bf{q}$ and $\bf{q}'$ are the wavevectors of the perturbing and induced density fluctuations, respectively.
If many-body effects can be neglected, this picture can be simplified for dipole-allowed transitions. In this case, using the spherical harmonics as basis for the target states and referring to transitions originating from a single state (as in $s-p$ excitations), the MDFF is \cite{Loffler201339}
\begin{widetext}
\begin{equation}
 S\left( \bf{q}, \bf{q'}; \it{E} \right) \propto \sum_{\mu L M, \mu' L' M'}{\left\langle j_{\lambda=1} \left( q \right) \right\rangle_{L E} \left\langle j_{\lambda=1} 
\left( q' \right)  \right\rangle_{L' E}}  
Y^{\mu}_{\lambda=1} \left( \bf{q} \right)^{*} Y^{\mu'}_{\lambda=1} \left( \bf{q'} \right)
 \Xi_{\lambda=1 \mu L M, \lambda=1 \mu' L' M'}\left( E \right)
\end{equation}
\end{widetext}
where $Y^{\mu}_{\lambda=1} \left( \bf{q} \right) $ are spherical harmonics, $ \left\langle j_{\lambda=1} \left( q \right) \right\rangle_{L E} $ is an integral of the spherical Bessel function $j_{\lambda=1}\left( q \right)$ weighted over the initial and final radial wavefunctions. $L$ and $M$ indicate the azimuthal and magnetic quantum number of the final state of the target electron, and $\lambda$ and $\mu$ are the angular momenta transferred during the transition. $\Xi_{ \lambda \mu L M,  \lambda' \mu' L M} \left( E \right)$ is a quantity that describes crystal-field effects and is proportional to the cross-density of states (XDOS)
\begin{equation}
\sum_{n \bf{k}}D^{n\bf{k}}_{LM} \left( D^{n\bf{k}}_{L'M'} \right)^{*} \delta \left( E_n \left( \bf{k} \right) -E \right)
\end{equation}
where $D^{n\bf{k}}_{LM}$ is the angular part of the final wave function, $n$ is the band index, and $\bf{k}$ is a k-point in the first Brillouin zone. Compared to the density of states (DOS), the XDOS includes also non-diagonal terms connecting states with different angular momenta. As $\Xi_{\lambda=1 \mu L M, \lambda=1 \mu' L M} \left( E \right)$ is a hermitian matrix \cite{Loffler201339}, the MDFF can be diagonalized. Therefore, assuming that the target's final states of the $s-p$ excitation are not degenerate, the transition matrix elements reflect the azimuthal shape of the final single-particle states and can thus be separated by using energy-filtering.

Ground-state calculations are performed using density-functional theory and the full-potential  augmented planewave plus local-orbital method, as implemented in \exciting \cite{exciting3}. Introducing a vacancy or a substitutional atom, a 5$\times$5 supercell is set up, hosting 49 and 50 atoms, respectively (Fig. 1 and 2 in the Supplemental Material \cite{supp}). The space group and thus the number of inequivalent carbon atoms (13) is the same in both cases. 
We adopt a lattice parameter of $a$=4.648 bohr, corresponding to a bond length of 2.683 bohr, while the cell size perpendicular to the graphene plane is set to c=37.794 bohr in order to prevent interactions between the periodically repeated layers. Exchange-correlation effects are treated by the PBE functional \cite{PhysRevLett.77.3865}.
The Brillouin zone is sampled with an 8$\times$8$\times$1 k-point grid. The structures are relaxed down to a residual force lower than 0.0005 Ha/bohr acting on each atom. Interatomic distances between atoms of the relaxed structures, up to the seventh nearest neighbor, are given in Table \ref{tab1}. Upon relaxation, the atoms surrounding the vacancy move slightly away from it, thus shortening the bond lengths with the next nearest neighbors, $d_{1-2}$, compared to the unperturbed system. The effect of the vacancy extends up to the fourth neighbours, whereas it is almost negligible for more distant atoms (more information about the relaxed structures can be found in the Supplemental Material \cite{supp}). In the case of nitrogen doping, the substitutional atom does not strongly influence the atomic configuration of the system. This happens because the nitrogen-carbon bond length is just slightly shortened with respect to the carbon-carbon bond length in pristine graphene. For all the systems, we have investigated dipole-allowed transitions at the K-edge of carbon, assuming an incoming electron beam perpendicular to the graphene plane. 
\begin{table}
\caption{\label{tab1} Bond lengths between atoms up to the seventh nearest neighbor, $d_{0-1}$, $d_{1-2}$, $d_{2-3}$, $d_{2-4}$, $d_{4-6}$, and $d_{6-7}$, for graphene doped with nitrogen (top row) and with a vacancy (bottom row). $\Delta d$ are the relative deviations from those of pristine graphene. $0$ indicates the defect site. (} 
\centering
\begin{tabular}{lc c c c c c c}
\hline \hline
System &  &  $d_{0-1}$ & $d_{1-2}$ & $d_{2-3}$& $d_{2-4}$ & $d_{4-6}$ & $d_{4-7}$\\
\hline\hline
N-doped & $d$  [bohr] & 2.675 & 2.675 & 2.683 & 2.689 & 2.683 & 2.692\\ 
& $\Delta d$ & -0.3\% & -0.3\% & - & +0.2\% & - & +0.3\%\\ [1ex]
\hline
Vacancy & $d$  [bohr] &  2.689 & 2.665 & 2.678 & 2.712 & 2.676 & 2.687\\
&  $\Delta d$ & +0.2\% & -0.7\%  & -0.2\% & +1.1\% & -0.3\% & +0.1\%\\  [1ex]
\hline
\end{tabular}
\end{table}

\begin{figure}[h]
 \begin{center}
 \bigskip
  \includegraphics[width=0.48\textwidth,angle=0]{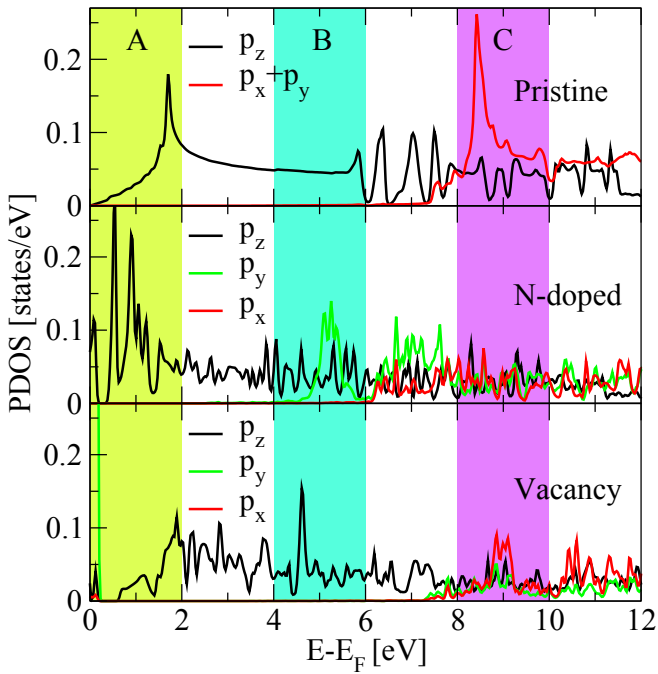}
\caption{\label{fig_DOS} Local projected density of states (PDOS) of carbon in pristine graphene (upper panel), and of the first nearest-neighbor atom for nitrogen-doped graphene (middle panel) and graphene with a single vacancy (bottom). $p_{x}$, $p_{y}$, and $p_{z}$ states are indicated by the red, green, and black lines, respectively. Note that, in the lower panel, the peak close to the Fermi energy exceeds the boundary of the box, with a maximum at about 2 states/eV. In the case of pristine graphene, the red line indicates the sum of $p_x$ and $p_y$. The three colored areas labeled as A, B, and C indicate the energy ranges for which the TEM images have been calculated.}
 \end{center}
\end{figure}

In Fig. \ref{fig_DOS}, the projected density of states (PDOS) of pristine graphene (upper panel) and of the first nearest-neighbor atom for nitrogen-doped graphene (middle panel) and graphene with a single vacancy (bottom), respectively, is plotted for empty states up to 12 eV above the Fermi energy. Here, $x$, $y$, and $z$ represent the local Cartesian coordinates at the individual atomic sites as determined by the point-group symmetry. In particular, $z$ is the axis perpendicular to the graphene, i.e., ($x$, $y$) plane. All the other atoms of the defective systems exhibit a PDOS with very similar character as in pristine graphene, besides the second and third nearest neighbors which are slightly affected by the defect \cite{supp}.
In pristine graphene, antibonding $\pi^*$ and $\sigma^*$ states are clearly recognizable at about 2 eV and 9 eV, respectively. As already reported in literature \cite{PhysRevB.76.115423, PhysRevB.85.165439, PhysRevB.68.144107,PhysRevB.86.045448}, the introduction of a vacancy or a substitutional nitrogen has a significant influence on the electronic structure. 
A consequence of the doping atom is lifting the degeneracy of $p_{x}$ and $p_{y}$ that is significant for the first nearest neighbors (middle panel in Fig. \ref{fig_DOS}). This effect is particularly evidenced by the appearance of bands at about 5 eV, which exhibit $p_{y}$ character. Here, three different regions can be easily identified: a) From 0 eV to 4 eV, the bands have only $p_{z}$ character; energy ranges that present such DOS character will be referred to as $T_{p_{z}}$. b) For energies higher than 6 eV, there are contributions from $p_{x}$, $p_{y}$, and $p_{z}$. The only difference to ideal graphene is the lifted degeneracy of $p_{x}$ and $p_{y}$. This defines a new kind of region, named $T_{p_{x,y,z}}$. c) Between 4 eV and 6 eV, the $p_{x}$ character of the first nearest neighbor is much less pronounced than that of $p_{y}$, while all the other atoms have only $p_{z}$ character; this region will be referred to as $T_{p_{y,z}}$.
In the case of graphene with a vacancy, the same kinds of regions can be identified, but corresponding to different energy ranges. Here, the $T_{p_z}$ type is found between 0.5 eV and 7 eV; $T_{p_{x,y,z}}$ encompasses energies above 7 eV; $T_{p_{y,z}}$ is a small energy window, just few tenths of eV close to the Fermi energy. We find similar kinds of DOS characters also for damaged nitrogen-doped graphene, i.e., graphene with a substitutional nitrogen and a vacancy located near it; such defects have beed recently reported in TEM measurements of nitrogen-doped graphene \cite{susi2012}. Details of this calculation and the corresponding simulated TEM images can be found in the Supplemental Material \cite{supp}.

\begin{figure}
 \centering
 \includegraphics[width=0.48\textwidth,angle=0]{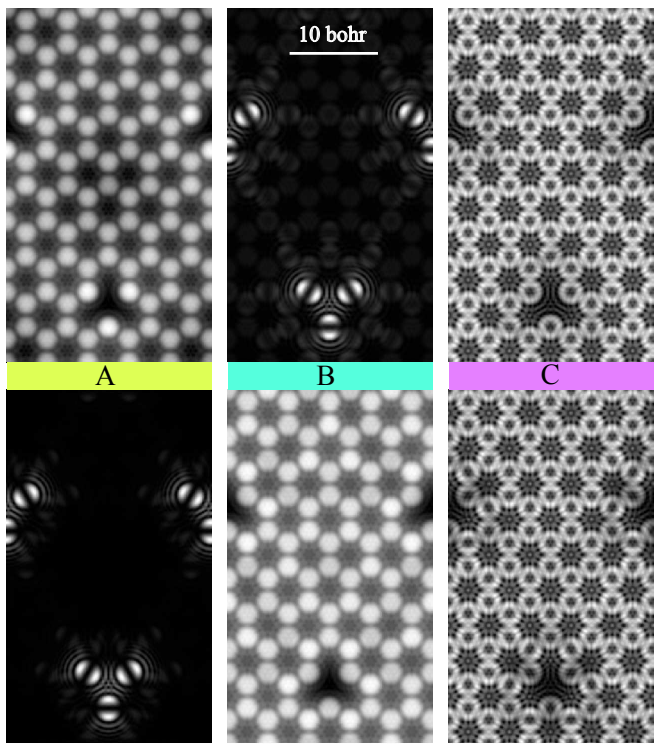}
 \caption{\label{fig_THEO} Simulated real-space intensity of the electron's exit wave function after propagation of an incident planewave through a graphene layer in presence of a nitrogen substitutional atom (upper panels) and a vacancy (bottom panels). The colored lines indicate the energy regions highlighted in Fig. \ref{fig_DOS}.}
\end{figure}

To investigate the impact of the local electronic structure (PDOS) on the EFTEM images, we first consider the ideal case of a perfect microscope with an acceleration voltage of 300 keV. In this case, the recorded images correspond to the intensity of the exit wavefunction in the multislice simulation \citep{Cowley}. The finite resolution of the spectrometer is taken into account by simulating images every 0.05 eV in 2 eV-broad energy ranges (regions A, B, and C in Fig. \ref{fig_THEO}) and then summing them up to get the final images. Each image is shown in contrast-optimized grayscale.

First, we analyze graphene with nitrogen doping (Fig. \ref{fig_THEO}, upper panels). Here, in the region close to the Fermi level (A in Fig. \ref{fig_DOS}) there are only contributions from $p_{z}$ orbitals. The image is then formed by disk-like features where their arrangement clearly visualizes the missing atoms (upper left panel in Fig. \ref{fig_THEO}). At an energy loss between 4 and 6 eV above the carbon K-edge (region B in Fig. \ref{fig_DOS}), there is a $T_{p_{y,z}}$-like region. We expect to see contributions from $p_{y}$ of the atom closest to the nitrogen, but no (or very little) signal coming from the other atoms. This happens because $p_{y}$ lies on a plane perpendicular to the incoming electron beam and its magnitude is more intense than the one of $p_{z}$; thus its contribution to the final signal overcomes the one from $p_{z}$ states. Consequently, only the $p_{y}$ orbitals of the three atoms surrounding the nitrogen are visible, which are pointing towards the defect, as imposed by the local $D_{3h}$ symmetry (upper middle panel in Fig. \ref{fig_THEO}). At an excitation energy between 8 and 10 eV above the K-edge (region C in Fig. \ref{fig_DOS}), instead, there are contributions from all the $p$ states. Since, however, $p_{x}$ and $p_{y}$ lie in a plane perpendicular to the beam axis, their contribution to the final signal dominates over the one from $p_{z}$ states. As a consequence, the image is composed of ring-like features, stemming solely from $p_{x}$ and $p_{y}$ states, arranged in hexagons (upper right panel in Fig. \ref{fig_THEO}). Due to symmetry breaking, the intensity is not uniform, neither along a ring (since $p_{x}$ and $p_{y}$ states are non-degenerate) nor among different rings (due to non-equivalent atomic sites).

The corresponding images for the system with a vacancy (bottom panels in Fig. \ref{fig_THEO}) appear nearly identical to the ones above, but at different energy ranges. This can be understood by comparing the PDOS of the two systems. Between 0 and 2 eV, for instance, we have a $T_{p_z}$-like region in the case of nitrogen-doped graphene, and both $T_{p_{y,z}}$ and $T_{p_z}$ in the case of graphene with a vacancy. Due to the similarity of the two systems, we will, in the following, focus on doped graphene and show the corresponding analysis for graphene with a single vacancy in the Supplemental Material \cite{supp}.

\begin{figure}[htbp]
 \centering
 \includegraphics[width=0.48\textwidth,angle=0]{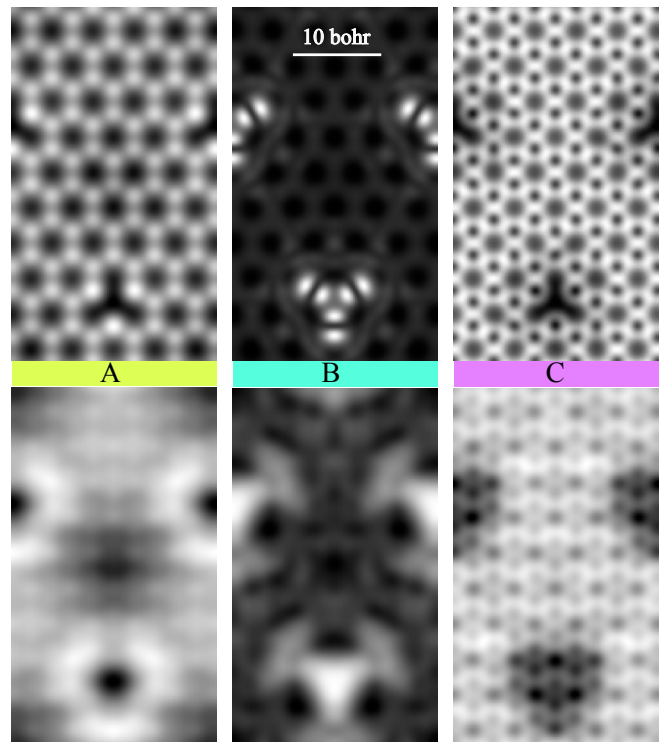}
 \caption{\label{fig_EXP_N} Simulated TEM image of nitrogen-doped graphene. The coloured lines indicate the energy regions highlighted in Fig. \ref{fig_DOS}. 80 keV incident beam energy and lenses as in a Titan (upper panels) and a Tecnai microscope (bottom panels) were assumed.}
\end{figure}

In order to predict the outcome of real experiments, we now visualize the effect of the optics' aberrations and of a more realistic acceleration voltage on these images. We have simulated an electron-beam acceleration voltage of 80 keV and the operating parameters of two different kinds of microscopes, the FEI Tecnai G$^2$ F20 and FEI Titan G$^2$ 60-300. The former has a spherical aberration $C_S$ = 1.2 mm, corresponding to a an extended Scherzer defocus of 849 $\textnormal{\AA}$, while the latter is a last-generation aberration-corrected microscope, i.e., exhibiting no spherical and chromatic aberrations. In view of that, chromatic aberrations are not included in the calculations. The images corresponding to the energy regions A, B, and C are shown in Fig. \ref{fig_EXP_N}.
Because of the lower resolution of the Tecnai microscope, all the features are blurred (lower panels) compared to the ideal images. Therefore, neither the atomic positions, nor the orbital shapes can be retrieved from them. On the other hand, images simulated by taking into account the aberration-corrected optics of the Titan microscope are very sharp and let us identify all the features already observed for the idealized situation previously described. This can be easily seen, comparing the upper panels of Figs. \ref{fig_EXP_N} and \ref{fig_THEO}. In particular, at an energy loss of 5 eV (region B), the $p_{y}$ orbitals are visible, as in the ideal images. This clearly demonstrates the potential ability of aberration-corrected microscopes to visualize atomic orbitals with EFTEM, especially in a system like graphene. This conclusion also holds when considering noise caused by the finite electron dose (see Supplemental Material \cite{supp} for corresponding images). 

In summary, we have predicted the possibility of performing orbital mapping in low-dimension systems using EFTEM and we have demonstrated it with the prototypical example of defective graphene. In particular, we have shown that, as far as the optics is concerned, reasonable image resolution may already nowadays be experimentally achievable with last generation aberration-corrected microscopes like a FEI Titan G$^2$ 60-300 and even more with improved instruments of the next generation. However, additional work is necessary to reduce artifacts such as noise, drift, instabilities and damage. The inelastic cross section for the carbon K-edge ionisation is about a factor of 10 smaller than the elastic scattering cross section on a carbon atom \cite{reimer}. The intensity collected within an energy window of 2 eV as in Fig. \ref{fig_THEO} is $<$5\% of the total K-edge intensity. So, order-of-magnitude-wise, in order to obtain the same SNR as in elastic imaging, we need at least 200 times more incident dose which means a dwell time of the order of several minutes for last generation TEMs in EFTEM mode. There is no fundamental law that would forbid such an experiment with today's  equipment; however, it is hampered by drift (which must be well below the interatomic distance during the exposure time), instabilities, and radiation damage. A new route to circumvent radiation damage based on an EFTEM low-dose technique was proposed recently \cite{Meyer201413}. This may solve the problem in future.

We have identified three different kinds of images that are expected to be acquired in an EFTEM experiment, depending on the character of the DOS: When only $p_{z}$ states are present in the electronic structure, the corresponding images are composed of disk-like features. When the DOS is characterized by contributions from all $p$ states, ring-like features are seen that, however, only originate from a convolution of $p_{x}$ and $p_{y}$ states, while the $p_z$ character is not visible. When the $p_{y}$ character strongly exceeds the one of $p_{x}$, only a single orbital is recorded. We expect this work to trigger new experiments on defective graphene and similar systems.

\section{Acknowledgments}
Financial support by the Austrian Science Fund (projects I543-N20 and J3732-N27) and the German Research Foundation within the DACH framework is acknowledged.


\section{Bibliography styles}
%
\end{document}